\documentclass[a4paper,12pt]{article}
\usepackage{amssymb}
\usepackage{amsmath}

\makeatletter \@addtoreset{equation}{section} \makeatother

\begin{document}


\begin{titlepage}

    \thispagestyle{empty}
    \begin{flushright}
        \hfill{CERN-PH-TH/2007-112} \\\hfill{UCLA/07/TEP/15}\\
    \end{flushright}

    \vspace{5pt}
    \begin{center}
        { \Large{\textbf{4d/5d Correspondence for the Black Hole Potential and its Critical Points}}}\vspace{25pt}
        \vspace{20pt}

        { {\textbf{{Anna Ceresole$^{\star}$, Sergio Ferrara$^{\diamondsuit\clubsuit\flat}$ and\ Alessio Marrani$^{\heartsuit\clubsuit}$}}}}\vspace{15pt}

        {$^\star$ \it INFN, Sezione di Torino and\\
        Dipartimento di Fisica Teorica, Universit\`{a} di Torino,\\
        Via Pietro Giuria 1, 10125 Torino, Italy\\
        \texttt{ceresole@to.infn.it}}

        \vspace{8pt}

        {$\diamondsuit$ \it Physics Department,Theory Unit, CERN, \\
        CH 1211, Geneva 23, Switzerland\\
        \texttt{sergio.ferrara@cern.ch}}

        \vspace{8pt}

        {$\clubsuit$ \it INFN - Laboratori Nazionali di Frascati, \\
        Via Enrico Fermi 40, 00044 Frascati, Italy\\
        \texttt{marrani@lnf.infn.it}}

        \vspace{8pt}

         {$\flat$ \it Department of Physics and Astronomy,\\
        University of California, Los Angeles, CA USA}

         \vspace{8pt}

        {$\heartsuit$ \it Museo Storico della Fisica e\\
        Centro Studi e Ricerche ``Enrico Fermi"\\
        Via Panisperna 89A, 00184 Roma, Italy}

        \vspace{15pt}
\end{center}


\begin{abstract}
We express the $d=4$, $\mathcal{N}=2$ black hole effective potential
for cubic holomorphic $F$ functions and generic dyonic charges in
terms of $d=5$ real special geometry data. The $4d$ critical points
are computed from the $5d$ ones, and their relation is elucidated.
For symmetric spaces, we identify the BPS and non-BPS classes of
attractors and the respective entropies. These always derive from
simple interpolating formul\ae\ between four and five dimensions,
depending on the volume modulus and on the $4d$ magnetic (or
electric) charges.

\end{abstract}

\end{titlepage}
\newpage

\section{Introduction\label{Intro}\protect\smallskip}

Recently there has been an increasing amount of work on extremal
charged black holes in an environment of scalar background fields,
as they naturally arise in modern theories of gravity: superstrings,
$M$-theory, and their low-energy description through supergravity.
In particular, the Attractor Mechanism for extremal black holes
\cite {FKS,Strom,FK1,FK2,FGK} in four and five dimensions has been
widely investigated for both $\mathcal{N}=2$ and extended
supergravities \cite {Moore}-\nocite
{Gaiotto-1,Gaiotto-2,Sen-old1,BCM,GIJT,Sen-old2,K1,TT,G,GJMT,Ebra1,K2,Ira1,Tom,BFM,
AoB-book,FKlast,Ebra2,BFGM1,rotating-attr,K3,Misra1,Lust2,BFMY,CdWMa,
DFT07-1,BFM-SIGRAV06,COP,Cer-Dal,ADFT-2,Saraikin-Vafa-1,Ferrara-Marrani-1,
TT2,ADOT-1,Ferrara-Marrani-2} \cite{CCDOP} (see also \cite{ADFT},
\cite{Larsen-rev} and \cite{Mohaupt-rev} for recent reviews).

The latest studies on BPS and non-BPS attractor points have
developed along two main lines.

In a top down approach \cite
{BFGM1,Ferrara-Marrani-1,Ferrara-Marrani-2,FG2,FG1} one uses some
powerful group theoretical techniques, descending from the geometric
properties of moduli spaces and $U$-duality invariants \cite
{ADFFT,ADF2,ADF,ADF-d=5,Ferrara-Gimon}, to classify the solutions to
the attractor equations and their properties. For theories with a
symmetric scalar manifold, these methods have led to a general
classification of BPS and non-BPS attractors, as well as studies of
their classical stability and entropy \cite
{BFGM1,Ferrara-Marrani-1,Ferrara-Marrani-2,FG2,FG1}. For $\mathcal{N}=2$, $%
d=4$ theories, some of these results have also been extended to more
general scalar manifolds based on cubic holomorphic prepotentials.
These so-called special K\"{a}hler ``$d$-geometries'' \cite{dWVVP}
are particularly relevant, as they naturally arise in the large
volume limit of Calabi-Yau compactifications of type IIA
superstrings. They include all special K\"{a}hler coset manifolds
$G/H$, that contain symmetric spaces are a further subclass.
Moreover, in $\mathcal{N}=2$, $d=4$ supegravity, cubic geometries
are precisely those that can be uplifted to five dimensions. Indeed,
all (rank-$3$) symmetric special K\"{a}hler spaces fall into this
class, and they admit a $d=5$ uplift to (rank-$2$) symmetric real
special spaces \cite{dWVVP,GST1,GST2,GST3,GST4,CVP,CKV-et-al}. We
remind that a generic $d$-geometry of complex dimension $n$ is not
necessarily a coset space, but nevertheless it has $n+1$ isometries,
corresponding to the shifts of the $n$ axions, and to an overall
rescaling of the prepotential (see \textit{e.g.} \cite{dWVVP} and
Refs. therein).

Conversely, in a bottom up approach, one attempts to construct
solutions (BH, magnetic strings, black rings, etc.) for a given
background spacetime geometry, by solving explicitly the equations
of motion \cite {FGK,Moore,Denef,TT,K3,BKRSW,Shmakova}. These are
originally the second order differential field equations for the
scalars and the warp factors, but they have been shown to be
equivalent to first order flow equations for both supersymmetric and
broad classes of non-supersymmetric, static and rotating BH
solutions \cite{Cer-Dal,ADOT-1,CCDOP}. In this context, the relation
between five and four dimensions is implemented through dimensional
reduction and by a Taub-NUT geometry for the black hole (see for
instance \cite{Gaiotto-1, Gaiotto-2,BCM,CCDOP,COP}).

This note brings closer these two lines of analysis and aims at
further exploiting, in the top down approach based on the $4d$
black-hole effective potential, the $5$-dimensional origin of
$\mathcal{N}=2$, special K\"{a}hler $d$-geometries. To this end, we
first write down the effective black hole potential in terms of the
5d real special geometry data, for generic dyonic charges and scalar
field values. Then we proceed to extremisation of this potential
with respect to the moduli, and we characterize the attractor points
in a 5d language. This will be achieved by connecting the critical
points of the BH effective potential through the interpolating
formul\ae\ between four and five dimensions. Furthermore, we  derive
the corresponding entropies which, for symmetric scalar manifolds,
are known to be given by the cubic and quartic invariants of the
$U$-duality groups, built solely in terms of the bare electric
and/or magnetic charges of the given BH configuration \cite
{ADFFT,ADF2,ADF,ADF-d=5,Ferrara-Gimon}. Notice that, compared to
previous literature such as \cite{BCM,CCDOP,COP}, most of our
formul\ae\ hold for generic points in moduli space, and$\backslash
$or that they cover both BPS and non-BPS solutions.\medskip

The paper is organized as follows. In Section \ref{Sect2} we recall
some
formul\ae\ and results holding for special $d$-geometries of $\mathcal{N}=2$%
, $d=4$ and $d=5$ supergravity. To this end, we work in the basis of
special coordinates, as they are those that naturally provide the
link to five dimensions. Thence, we compute the effective potential
$V$ for generic dyonic charges and $d$-geometry, and we give its
properties for specific BH charge configuration where it undergoes
some simplifications. In Section \ref {Sect3} we give $4d$
attractors in terms of $5d$ ones, and we compute the corresponding
BH entropy. For symmetric spaces, the interpolation yields a
clear relation between the $\mathcal{N}=2$, $d=4$ and $d=5$ $\frac{1}{2}$%
-BPS and non-BPS BH charge orbits studied in the literature \cite
{BFGM1,Ferrara-Marrani-1,Ferrara-Marrani-2,FG2,FG1}, which is
developed in Section \ref{Sect4}. Finally, some further comments and
outlooks are given in Section \ref{Conclusion}.

\section{Special Geometry for Cubic Holomorphic Prepotentials}

\label{Sect2}

We consider $\mathcal{N}=2$, $d=4$ special K\"{a}hler geometry based, in
special coordinates $X^{\Lambda }=\left( X^{0},X^{0}z^{i}\right) $, on the
holomorphic prepotential
\begin{equation}
\begin{array}{l}
F\left( X\right) =\frac{1}{3!}d_{ijk}\frac{X^{i}X^{j}X^{k}}{X^{0}}=\left(
X^{0}\right) ^{2}f\left( z\right) ; \\
\\
f\left( z\right) \equiv \frac{1}{3!}d_{ijk}z^{i}z^{j}z^{k}. \label{effe}
\end{array}
\end{equation}
In the K\"{a}hler gauge $X^{0}\equiv 1$ and in the special coordinate basis,
the K\"{a}hler potential reads ($f_{i}\equiv \frac{\partial f\left( z\right)
}{\partial z^{i}}=\frac{1}{2}d_{ijk}z^{j}z^{k}$)
\begin{equation}
\begin{array}{l}
K=-ln\left( Y\right) ; \\
\\
Y\equiv i\left[ 2(f-\overline{f})+(\overline{z}^{\overline{\imath }%
}-z^{i})(f_{i}+\overline{f}_{\overline{\imath }})\right] =-\frac{i}{3!}%
d_{ijk}(z^{i}-\overline{z}^{\overline{\imath }})(z^{j}-\overline{z}^{%
\overline{\jmath }})(z^{k}-\overline{z}^{\overline{k}}).
\end{array}
\end{equation}
By defining the real components of the $d=4$ moduli as $z^{i}=x^{i}-i\lambda
^{i}$, one gets for the K\"{a}hler potential
\begin{equation}
K=-ln\left( 8\mathcal{V}\right) ,~~~\mathcal{V}\equiv \frac{1}{3!}%
d_{ijk}\lambda ^{i}\lambda ^{j}\lambda ^{k},
\end{equation}
and therefore also the K\"{a}hler metric $g_{i\bar{\jmath}}\equiv \partial
_{i}\partial _{\bar{\jmath}}K$ becomes a real function of the $\lambda ^{i}$
variables only \cite{Vaula},
\begin{equation}
\begin{array}{l}
g_{i\bar{\jmath}}\equiv g_{ij}=-\frac{3}{2}\left( \frac{\kappa _{ij}}{\kappa
}-\frac{3}{2}\frac{\kappa _{i}\kappa _{j}}{\kappa ^{2}}\right) =-\frac{1}{4}%
\frac{\partial ^{2}ln\left( \mathcal{V}\right) }{\partial \lambda
^{j}\partial \lambda ^{i}}\Leftrightarrow g^{i\bar{\jmath}}=2\left( \lambda
^{i}\lambda ^{j}-\frac{\kappa }{3}\kappa ^{ij}\right) \equiv g^{ij}; \\
\\
\kappa _{ij}\equiv d_{ijk}\lambda ^{k},~~\kappa _{i}\equiv d_{ijk}\lambda
^{j}\lambda ^{k},~~\kappa \equiv d_{ijk}\lambda ^{i}\lambda ^{j}\lambda
^{k}=6\mathcal{V},~~\kappa ^{ij}\kappa _{jl}\equiv \delta _{l}^{i};
\end{array}
\end{equation}
We introduce the $d=5$ real moduli as $\lambda ^{i}\equiv \mathcal{V}^{1/3}%
\widehat{\lambda }^{i}$. They satisfy
\begin{equation}
\frac{1}{3!}d_{ijk}\widehat{\lambda }^{i}\widehat{\lambda }^{j}\widehat{%
\lambda }^{k}=1\, ,
\end{equation}
which is the defining equation of the $d=5$ real special manifold. In these
variables one gets
\begin{equation}
\begin{array}{l}
g_{ij}=\frac{1}{4}\left( \frac{1}{4}\widehat{\kappa }_{i}\widehat{\kappa }%
_{j}-\widehat{\kappa }_{ij}\right) \mathcal{V}^{-2/3}=\frac{1}{4}\mathcal{V}%
^{-2/3}a_{ij}\Leftrightarrow g^{ij}=2\left( \widehat{\lambda }^{i}\widehat{%
\lambda }^{j}-2\widehat{\kappa }^{ij}\right) \mathcal{V}^{2/3}=4\mathcal{V}%
^{2/3}a^{ij}; \\
\\
\widehat{\kappa }_{ij}\equiv d_{ijk}\widehat{\lambda }^{k},~~\widehat{\kappa
}_{i}\equiv d_{ijk}\widehat{\lambda }^{j}\widehat{\lambda }^{k},~~\widehat{%
\kappa }\equiv d_{ijk}\widehat{\lambda }^{i}\widehat{\lambda }^{j}\widehat{%
\lambda }^{k}=6,~~\widehat{\kappa }^{ij}\widehat{\kappa }_{jl}\equiv \delta
_{l}^{i},
\end{array}
\end{equation}
where
\begin{equation}
a_{ij}\equiv \left. 4g_{ij}\right| _{\mathcal{V}=1},~~a^{ij}a_{jk}=\delta
_{k}^{i}\Longrightarrow a_{ij}\widehat{\lambda }^{j}=\frac{1}{2}\widehat{%
\kappa }_{i},~a_{ij}\widehat{\lambda }^{i}~\widehat{\lambda }^{j}=3.
\end{equation}
One can then proceed to computing also the vector kinetic matrix $\mathcal{N}%
_{\Lambda \Sigma }$ in terms of these quantities, obtaining
\begin{equation}
\begin{array}{l}
Im\mathcal{N}_{\Lambda \Sigma }=-\mathcal{V}\left(
\begin{array}{ccc}
1+4g &  & g_{j} \\
&  &  \\
g_{i} &  & 4g_{ij}
\end{array}
\right) \Leftrightarrow \left( Im\mathcal{N}_{\Lambda \Sigma }\right) ^{-1}=-%
\frac{1}{\mathcal{V}}\left(
\begin{array}{ccc}
1 &  & x^{j} \\
&  &  \\
x^{i} &  & x^{i}x^{j}+\frac{1}{4}g^{ij}
\end{array}
\right) ; \\
\\
Re\mathcal{N}_{\Lambda \Sigma }=\left(
\begin{array}{ccc}
\frac{1}{3}h &  & -\frac{1}{2}h_{j} \\
&  &  \\
-\frac{1}{2}h_{i} &  & h_{ij}
\end{array}
\right) ,
\end{array}
\label{ImN-ReN-d-SKG}
\end{equation}
where
\begin{equation}
\begin{array}{l}
g\equiv g_{ij}x^{i}x^{j},~~g_{i}\equiv -4g_{ij}x^{j}; \\
\\
h_{ij}\equiv d_{ijk}x^{k},~~h_{i}\equiv d_{ijk}x^{j}x^{k},~~h\equiv
d_{ijk}x^{i}x^{j}x^{k}.
\end{array}
\end{equation}
Note that in the above special coordinate basis, the index $0$, associated
to the graviphoton vector, is naturally split from the $d=5$ index $%
i=1,...,n_{V}$. In this language, the $d=5$ real special manifold is simply
the $\left( n_{V}-1\right) $ real hypersurface with unit volume
. Accordingly, the $n_{V}$ $d=4$ complex moduli $z^{i}$ separate into $%
\left( x^{i},\mathcal{V},\widehat{\lambda }^{i}\right) $, where $\widehat{%
\lambda }^{i}$ are the $n_{V}$ real positive $d=5$ moduli, parameterizing
the hypersurface $
\frac{1}{3!}d_{ijk}\widehat{\lambda }^{i}\widehat{\lambda }^{j}\widehat{%
\lambda }^{k}=1$.

Let us now consider the $d=4$ BH effective potential,
\begin{equation}
V=-\frac{1}{2}Q^{T}\mathcal{M}Q,  \label{VBH-d=4}
\end{equation}
where $Q$ denotes the $\left( 2n_{V}+2\right)$ charge vector
\begin{equation}
Q=\left( p^{0},p^{i},q_{0},q_{i}\right) ,  \label{Q}
\end{equation}
and $\mathcal{M}$ is the real symplectic $\left( 2n_{V}+2\right) \times
\left( 2n_{V}+2\right) $ matrix
\begin{equation}
\mathcal{M}\equiv \left(
\begin{array}{cccc}
Im\mathcal{N}+Re\mathcal{N}\left( Im\mathcal{N}\right) ^{-1}Re\mathcal{N} &
&  & -Re\mathcal{N}\left( Im\mathcal{N}\right) ^{-1} \\
&  &  &  \\
-\left( Im\mathcal{N}\right) ^{-1}Re\mathcal{N} &  &  & \left( Im\mathcal{N}%
\right) ^{-1}
\end{array}
\right) .  \label{M}
\end{equation}
By using Eqs. (\ref{VBH-d=4})-(\ref{M}) and the expressions computed in Eq. (%
\ref{ImN-ReN-d-SKG}), one obtains the following formula of the $d=4$
effective BH potential for a generic special K\"{a}hler $d$-geometry and
dyonic charges:
\begin{eqnarray}
&&
\begin{array}{l}
2V=\left[ \frac{\kappa }{6}\left( 1+4g\right) +\frac{h^{2}}{6\kappa }+\frac{3%
}{8\kappa }g^{ij}h_{i}h_{j}\right] \left( p^{0}\right) ^{2}+ \\
\\
+\left[ \frac{2}{3}\kappa g_{ij}+\frac{3}{2\kappa }\left(
h_{i}h_{j}+h_{im}g^{mn}h_{nj}\right) \right] p^{i}p^{j}+ \\
\\
+\frac{6}{\kappa }\left[ \left( q_{0}\right) ^{2}+2 x^{i} q_{0}q_{i}+\left(
x^{i}x^{j}+\frac{1}{4}g^{ij}\right) q_{i}q_{j}\right] + \\
\\
+2\left[ \frac{\kappa }{6}g_{i}-\frac{h}{2\kappa }h_{i}-\frac{3}{4\kappa }%
g^{jm}h_{m}h_{ij}\right] p^{0}p^{i}+ \\
\\
-\frac{2}{\kappa }\left[ -hp^{0}q_{0}+3q_{0}p^{i}h_{i}-\left( hx^{i}+\frac{3%
}{4}g^{ij}h_{j}\right) p^{0}q_{i}+3\left( h_{j}x^{i}+\frac{1}{2}%
g^{im}h_{mj}\right) q_{i}p^{j}\right] .
\end{array}
\notag \\
&&  \label{VBH-d=4-d-SKG}
\end{eqnarray}
A quick look reveals that the axions $x^{i}$ appear in $V$ through a
polynomial of degree $6$, whose coefficients depend on $\lambda ^{i}$ and on
$d_{ijk}$. Moreover, terms linear in $x^{i}$ vanish if $q_{0}q_{i}$, $%
p^{0}p^{i}$ and $p^{i}q_{j}$ separately vanish. This means that for BH
charge configurations of the type
\begin{equation}
\begin{array}{l}
a)~~Q_{0}=\left( p^{0},0,q_{0},0\right) ; \\
\\
b)~~Q_{e}=\left( p^{0},0,0,q_{i}\right) ; \\
\\
c)~~Q_{m}=\left( 0,p^{i},q_{0},0\right) . \label{cases}
\end{array}
\end{equation}
the Attractor Equations $\frac{\partial V}{\partial x^{i}}=0$ admit the $n$
solutions given by $x^{i}=0$, \textit{i.e.} by purely imaginary critical
moduli $z^{i}=-i\lambda ^{i}$.

\subsection{Symmetric $d$-geometries}

Something more can be said if one considers symmetric space theories
associated to scalar manifolds $G/H$, such that G is a symmetry of the
action. The discussion below will rely on the results of \cite{FG1,FG2,BFGM1}%
. All symmetric special K\"{a}hler $d $-geometries are known to originate
from five dimensional theories through dimensional reduction . In this case,
the tensor $d_{ijk}$ is an invariant tensor of the $U$-duality $d=5$ group $%
G_{5}$, and $\lambda ^{i}$ (and $x^{i}$) transform linearly under $G_{5}$
(we recall that $G_{4}$ decomposes into $G_{5}\otimes SO\left( 1,1\right) $,
where $SO\left( 1,1\right) $ corresponds to the radius of compactifications
along $S^{1}$ \cite{GST2}). Furthermore, the symmetric tensor $d_{ijk}$
satisfies the non-linear relation \cite{GST2,CVP}
\begin{equation}
d_{r(pq}d_{ij)k}d^{rkl}=\frac{4}{3}\delta _{(p}^{l}d_{qij)},
\label{symmetric-cond}
\end{equation}
which is equivalent to the condition for the corresponding manifold (in $d=4$
and $d=5$) to be symmetric.

For a symmetric real special manifold $\frac{G_{5}}{H_{5}}$, one can always
perform a suitable $H_{5}$-transformation that brings the cubic polynomial
to \textit{normal form},
\begin{equation}
I_3(q)=\frac{1}{3!}d^{ijk}q_{i}q_{j}q_{k}=q_{1}q_{2}q_{3},  \label{normal}
\end{equation}
where $q_{1}$, $q_{2}$ and $q_{3}$ are the three eigenvalues of the
corresponding $3\times 3$ Jordan system. For non-symmetric $d$-geometries
Eq. (\ref{normal}) does not hold any more, but nevertheless we will confine
our discussion of $d=5$ to the $3$-charge case. The simplest example of this
kind is provided by the $5$-dimensional uplift of the $stu$ model\cite
{Duff-stu,BKRSW,K3}, consisting in a $2$-dimensional free $\sigma $-model $%
SO(1,1)\otimes SO(1,1)$, whose real special geometry is determined by the
constraint
\begin{equation}
\frac{1}{3!}d_{ijk}\widehat{\lambda }^{i}\widehat{\lambda }^{j}\widehat{%
\lambda }^{k}=\widehat{\lambda }^{1}\widehat{\lambda }^{2}\widehat{\lambda }%
^{3}=1.
\end{equation}
This is the model we will use to perform most of the computations, even
though our results will hold in general for rank-$2$ symmetric real special
manifolds.

We now consider more in depth the BH charge configurations in (\ref{cases})
for symmetric $d$-geometries. In this case, the quartic invariant is given
by (see \cite{Ferrara-Gimon} for notation and further elucidation)
\begin{equation}
I_{4}\left( p^{0},p^{i},q_{0},q_{i}\right) =-\left(
p^{0}q_{0}+p^{i}q_{i}\right) ^{2}+4\left[ q_{0}I_{3}\left( p\right)
-p^{0}I_{3}\left( q\right) +\left\{ \frac{\partial I_{3}\left( p\right) }{%
\partial p},\frac{\partial I_{3}\left( q\right) }{\partial q}\right\} \right]
\,  \label{I4}
\end{equation}
in terms of the cubic invariants of the five-dimensional U-duality group $%
G_{5}$, with
\begin{equation}
I_{3}(p)=\frac{1}{3!}d_{ijk}p^{i}p^{j}p^{k}\,\,\,\,\,\,\,,%
\{I_{3}(q),I_{3}(p)\}\equiv \frac{\partial I_{3}(q)}{\partial q_{i}}\frac{%
\partial I_{3}(p)}{\partial p^{i}}\,\,\,.
\end{equation}
Note that all terms of Eq. (\ref{I4}) are invariant under $G_{5}$, because $%
p^{i}$s and $q_{i}$s transform as the linear gradient and contragradient
representation of $G_{5}$, respectively.

According to the classification of charge orbits for symmetric $d$%
-geometries, it is known that for $d=5$ there are two distinct orbits (one
BPS and the other non-BPS) \cite{FG1,FG2}, whereas for $d=4$ there exist
three orbits, one BPS and two non-BPS \cite{BFGM1}. For a given BH charge
configuration, the $d=5$ and $d=4$ charge orbits, respectively with $3$ and $%
4$ distinct eigenvalues, can actually cover all cases, as follows \cite
{FG1,FG2,BFGM1}:
\begin{equation}
\begin{array}{l}
d=5:\left\{
\begin{array}{l}
\text{BPS}:(+++)~\text{or~}(---)\,; \\
\\
\text{non-BPS}:(++-)~\text{or~}(--+)\,;
\end{array}
\right. ~~ \\
\\
\\
d=4:\left\{
\begin{array}{l}
\text{BPS}:(++++)~\text{or~}(----)\,; \\
\\
\text{non-BPS, }Z\neq 0:(+++-)~\text{or~}(---+)\,; \\
\\
\text{non-BPS, }Z=0:(++--)\,.
\end{array}
\right.
\end{array}
\end{equation}

For the BH charge configurations in (\ref{cases}), one gets
\begin{equation}
\begin{array}{l}
a)\, \, \, I_{4}\left( p^{0},q_{0}\right) =-\left( p^{0}q_{0}\right) ^{2};
\\
\\
b)\, \, \, I_{4}\left( p^{0},q_{i}\right) =-4p^{0}I_{3}\left( q\right); \\
\\
c)\, \, \, I_{4}\left( p^{i},q_{0}\right) =4q_{0}I_{3}\left( p\right) .
\label{orbite}
\end{array}
\end{equation}
While the configuration a), $\left( p^{0},q_{0}\right) $, clearly gives only
$I_{4}<0$, the other two configurations b) and c), which are reciprocally
dual, yield $I_{4}\gtrless 0$, depending on whether the cubic invariant $%
I_{3}\left( q\right) $ and $I_{3}\left( p\right) $ have the same or opposite
sign of $p^{0}$ or $q_{0}$. In the subsequent treatment of symmetric spaces
we will consider only one of these charge configurations, namely the \textit{%
electric} one associated to $Q_e$.

Let us take the $d=5$ ($\frac{1}{2}$-)BPS configuration where $q_{1}$, $%
q_{2} $, $q_{3}$ all have positive sign , so that $I_{3}\left(q\right) >0$ .
It is obvious that, depending on the sign of $p^{0}$ one will get a $d=4$ ($%
\frac{1}{2}$-)BPS or non-BPS $Z\neq 0$ configuration. Indeed, since $%
I_{3}\left( q\right) >0$ implies that $I_{4}\left( p^{0},q_{i}\right)
\gtrless 0$ according to $p^{0}\lessgtr 0$, one gets that $q_{1}$, $q_{2}$, $%
q_{3}>0$ is the $d=4$ BPS configuration, where we identify $\left(
-p^{0},q_{1},q_{2},q_{3}\right) $ with the $4$ positive eigenvalues of the
corresponding Freudenthal triple system.

On the other hand, if one starts with a $d=5$ non-BPS configuration, say
with $q_{1}$, $q_{2}<0$ and $q_{3}>0$, then, for either sign of $p^{0}$, one
always obtains a $d=4$ non-BPS configuration. The sign of $p^{0}$ is however
crucial, because it gives rise to two inequivalent non-BPS charge
configurations, distinguished by the vanishing of the central charge:
\begin{equation}
\begin{array}{l}
p^{0}<0\Rightarrow \text{non-BPS,~~}Z=0,~~I_{4}>0; \\
\\
p^{0}>0\Rightarrow \text{non-BPS,~~}Z\neq 0,~~I_{4}<0.
\end{array}
\end{equation}

In the next Section we will study this phenomenon by directly solving the
Attractor Equations.

\section{$5d$ and $4d$ Black Hole Potentials, Entropies and Their
Interpolation}

\label{Sect3}

In this Section we study the relation between the critical points of $%
V_{5}^{q}$ and their $d=4$ counterparts, for the BH charge configurations in
(\ref{cases}). More specifically, we will discuss the interpolation between $%
d=5$ and $d=4$ extremal BHs in $\mathcal{N}=2$ supergravity for some
particular cases, namely for the aforementioned BH charge configurations in (%
\ref{cases}). 
For $CY_{3}$-compactifications these charge configurations respectively
correspond to switching on the charges of $a)$ $D0-D6$ branes; $b)$ $D0-D4$
branes; $c)$ $D2-D6$ branes. As previously mentioned, configuration $b)$ is
called \textit{electric}, and its uplift to $d=5$ describes the so-called
extremal Tangherlini BH \cite{Tangh-1,Tangh-2} with horizon geometry $%
AdS_{2}\times S^{3}$. On the other hand, configuration $c)$ is called
\textit{magnetic}, and its uplift to $d=5$ describes a black string with
horizon geometry $AdS_{3}\times S^{2}$. For symmetric $d$-geometries
configurations $b)$ and $c)$ are reciprocally dual, but this does not hold
any more for a generic $d$-geometry. We will consider only configuration $b)$
in our treatment, briefly commenting at the end of the Section on the
extension of our results to non-symmetric $d$-geometries.


\subsection{$d=4$ Black Hole Potential for Vanishing Axions}

As previously noticed, the $\mathcal{N}=2$, $d=4$ extremal BH potential $V$,
given for a generic special K\"{a}hler $d$-geometry by Eq. (\ref
{VBH-d=4-d-SKG}), undergoes a dramatic simplification when no linear terms
in the axions $x^{i}$ appear, such that the criticality conditions $\frac{%
\partial V}{\partial x^{i}}=0$ can be solved by putting $x^{i}=0$ $\forall i$%
. For this situation, we will introduce the notation $V^{\ast}=V|
_{x^{i}=0~\forall i}$. \medskip

By further considering $Q_0=\left( p^{0},0,q_{0},0\right) $ (BH charge
configuration $a)$), Eq. (\ref{VBH-d=4-d-SKG}) yields 
\begin{equation}
V^{\ast} =\frac{1}{2}\left[ \left( p^{0}\right)^{2}\mathcal{V}+\left(
q_{0}\right) ^{2}\mathcal{V}^{-1}\right] = V^{\ast} \left( \mathcal{V}%
,p^{0},q_{0}\right) ,
\end{equation}
where we have used the relation\ $\kappa =6\mathcal{V}$. By recalling the
redefinition $\lambda ^{i}\equiv \mathcal{V}^{1/3}\widehat{\lambda }^{i}$,
it is immediate to realize that for the BH charge configuration $a)$ the
effective BH potential $V^{\ast}$ at vanishing axions \textit{does not
depend on any of the }$\widehat{\lambda }^{i}$, and thus it has $n_{V}-1$
``flat'' directions at all orders.

This result agrees with the findings of \cite{TT2}, and also with the
analysis performed in \cite{Ferrara-Marrani-1} and \cite{Ferrara-Marrani-2}.
Indeed, for symmetric spaces, the moduli space of the $d=4$ non-BPS $Z\neq 0$
orbit ($I_{4}<0$) given in \ref{orbite} coincides with the special real
scalar manifold of the $d=5 $ parent theory \cite{Ferrara-Marrani-2}.

It is easy to realize that this result actually holds for a generic $d$%
-geometry rather than for only symmetric ones. From a $d=5$ perspective, no $%
5d$ charges are turned on, because $p^{0}$ and $q_{0}$ are the charges of
the Kaluza-Klein vector, and thus one would not expect that the $5d$ moduli $%
\widehat{\lambda }^{i}$ are stabilized, as it actually happens.

Since
\begin{equation}
\frac{\partial V^{\ast}}{\partial \mathcal{V}}=0\Leftrightarrow \mathcal{V}%
=\left| \frac{q_{0}}{p^{0}}\right| ,  \label{p0-q0-crit-point}
\end{equation}
by defining
\begin{equation}
V^{\ast}_{\left( p^{0},q_{0}\right) ,crit.}\equiv \left. V^{\ast}\right| _{%
\frac{\partial V}{\partial \mathcal{V}}=0},
\end{equation}
one gets
\begin{equation}
\frac{S_{BH}}{\pi }=V^{\ast}_{\left( p^{0},q_{0}\right) ,crit.}=\left|
p^{0}q_{0}\right| .  \label{p0-q0-crit-V}
\end{equation}
It is interesting to observe that Eq. (\ref{p0-q0-crit-point}) and (\ref
{p0-q0-crit-V})are the same ones of the so-called \textit{dilatonic BH} (see
\cite{FK1} and Refs. therein). However, whereas the dilatonic BH is BPS, the
present case has $I_{4}<0$, and thus it corresponds to a non-BPS $Z\neq 0$
attractor.\medskip

Let us now move to consider $Q_{e}=\left( p^{0},0,0,q_{i}\right) $ (BH
charge configuration $b$), that is in $d=5$ a generic charge configuration
for the extremal Tangherlini BH. Then Eq. (\ref{VBH-d=4-d-SKG}) yields
\begin{equation}
V^{\ast }=\frac{1}{2}\left[ \left( p^{0}\right) ^{2}\mathcal{V}+\mathcal{V}%
^{-1/3}a^{ij}q_{i}q_{j}\right] =V^{\ast }\left( \mathcal{V},\widehat{\lambda
}^{i},p^{0},q_{j}\right) .  \label{27june-13}
\end{equation}
By defining the $5d$ black hole potential
\begin{equation}
V_{5}^{q}\equiv a^{ij}q_{i}q_{j}=V_{5}^{q}\left( \widehat{\lambda }%
^{i},q_{j}\right) ,  \label{27june-14}
\end{equation}
one obtains that
\begin{equation}
\frac{\partial V^{\ast }}{\partial \widehat{\lambda }^{i}}=0\Longrightarrow
\frac{\partial V_{5}^{q}}{\partial \widehat{\lambda }^{i}}=0,~\forall i.
\end{equation}
Since
\begin{equation}
\frac{\partial V^{\ast }}{\partial \mathcal{V}}=0\Leftrightarrow \mathcal{V}%
^{4/3}=\frac{1}{3\left( p^{0}\right) ^{2}}V_{5}^{q}~,  \label{27june-7}
\end{equation}
by defining
\begin{equation}
V_{\left( p^{0},q_{i}\right) ,crit.}^{\ast }\equiv V^{\ast }~~\text{at~~}%
\left\{
\begin{array}{l}
\frac{\partial V^{\ast }}{\partial \mathcal{V}}=0; \\
\\
\frac{\partial V^{\ast }}{\partial \widehat{\lambda }^{i}}=0~\forall i,
\end{array}
\right.
\end{equation}
one gets
\begin{equation}
\frac{S_{BH}}{\pi }=V_{\left( p^{0},q_{i}\right) ,crit.}^{\ast }=2\left|
p^{0}\right| ^{1/2}\left. \left( \frac{V_{5}^{q}}{3}\right) ^{3/4}\right| _{%
\frac{\partial V_{5}^{q}}{\partial \widehat{\lambda }^{i}}=0~\forall i},
\label{27june-1}
\end{equation}
a formula valid for any $d$-geometry (for vanishing axions, and in the BH
charge configuration $Q=\left( p^{0},0,0,q_{i}\right) $).

For a symmetric $d$-geometry, it holds that
\begin{equation}
\left. \frac{V_{5}^{q}}{3}\right| _{\frac{\partial V_{5}^{q}}{\partial
\widehat{\lambda }^{i}}=0~\forall i}=\left( \frac{1}{3!}\left|
d^{ijk}q_{i}q_{j}q_{k}\right| \right) ^{2/3}=\left| q_{1}q_{2}q_{3}\right|
^{2/3},  \label{27june-4}
\end{equation}
so one finally gets
\begin{equation}
\mathcal{V}_{crit.}^{4/3}=\frac{1}{\left( p^{0}\right) ^{2}}\left( \frac{1}{%
3!}\left| d^{ijk}q_{i}q_{j}q_{k}\right| \right) ^{2/3}=\frac{\left|
q_{1}q_{2}q_{3}\right| ^{2/3}}{\left( p^{0}\right) ^{2}}~,  \label{29june-3}
\end{equation}
and
\begin{equation}
\frac{S_{BH}}{\pi }=V_{\left( p^{0},q_{i}\right) ,crit.}^{\ast }=2\left|
p^{0}q_{1}q_{2}q_{3}\right| ^{1/2},  \label{27june-2}
\end{equation}
which is the known $d=4$ result \cite{BKRSW,K3,TT2}. Since the $d=5$ BH
entropy has the general form (for the electric charge configuration) (\cite
{FK1,FG2}; see also Eq. (\ref{VBH5-V5q}))
\begin{equation}
\frac{S_{BH,d=5}}{\pi }=\left( 3\left. V_{5}^{q}\right| _{\frac{\partial
V_{5}^{q}}{\partial \widehat{\lambda }^{i}}=0~\forall i}\right) ^{3/4},
\end{equation}
one also obtains that (see also Eq. (\ref{VBH5-crit}))
\begin{equation}
\frac{S_{BH,d=5}}{\pi }=3^{\frac{3}{2}}\left| q_{1}q_{2}q_{3}\right| ^{1/2}.
\label{SBH5}
\end{equation}
The above formulae may be compared to those obtained in \cite{COP} with a
different approach, that is in the context of the entropy function formalism
(which is known to hold only on shell for the scalar fields), and where
vanishing axions are those associated to non-rotating black holes (see also
\cite{CCDOP}).

We also remark that the use of a five-dimensional BH metric Ansatz implies
that, in order for the $4d$ BH entropy to have the correct form, the $5d$
charges must be redefined quadratically in terms of the $4d$ ones (for our
case $Q_{i}=-p^{0}q_{i}$, where $Q_{i}$ denotes the $5$-dimensional charges;
see \textit{e.g.} \cite{Pioline-rev}). In our derivation such redefinition
does not occur, because our approach directly takes into account the
symmetries of the problem\footnote{%
Let us also notice that in our notation $p^{0}$ and $q_{0}$ are opposite to
those used in \cite{Pioline-rev}.}.

Finally, the above computations and results are insensitive to the sign of
the BH charges. This fact can be understood by noticing that the charges
appear quadratically in $V^{\ast}$ specified for the background charge
vector $Q_{e}=\left( p^{0},0,0,q_{i}\right) $. On the other hand, the
supersymmetric nature of the solutions (\ref{27june-1}) and (\ref{27june-2})
crucially depend on the sign of the four eigenvalues $\left(
-p^{0},q_{1},q_{2},q_{3}\right) $ of the corresponding Freudenthal triple
system (see \textit{e.g.} \cite{BFGM1} and Refs. therein). In the next
Section, in the framework of the ($d=5$ uplift of the) $stu$ model, we will
see that, depending on the signs of the charges, the attractor configuration
determining the entropy (\ref{27june-2}) has the following
supersymmetry-preserving features:
\begin{equation}
\begin{array}{l}
I_{4}<0:~\text{non-BPS,~}Z\neq 0; \\
\\
I_{4}>0:~\text{\textit{either}~}\frac{1}{2}\text{-BPS~\textit{or}~non-BPS,~}%
Z=0.
\end{array}
\end{equation}

\section{BPS and non-BPS $d=5$ and $d=4$ Relations}

\label{Sect4}

We now turn to exploring the BPS/non-BPS nature of the 5d/4d attractors by
considering the simplest rank-$3$ special K\"{a}hler geometry, that is the $%
stu$ model. However, our results clearly hold for all rank-$3$ symmetric $d$%
-geometries, indeed all containing the $stu$ model. Thus, we need to
consider the homogeneous symmetric manifold $\left( \frac{SU\left(
1,1\right) }{U\left( 1\right) }\right) ^{3}$, product of three rank-$1$
cosets. The $d=5$ corresponding real special homogeneous symmetric manifold
is $SO(1,1)\otimes SO(1,1)=\mathbb{R}_{0}^{+}\times \mathbb{R}_{0}^{+}$,
\textit{i.e.} the rank-$2$ product of two rank-$1$ free trivial spaces. Such
manifold can be also characterized as the geometrical \textit{locus} ($%
i=1,2,3$)
\begin{equation}
\frac{1}{3!}d_{ijk}\widehat{\lambda }^{i}\widehat{\lambda }^{j}\widehat{%
\lambda }^{k}=\widehat{\lambda }^{1}\widehat{\lambda }^{2}\widehat{\lambda }%
^{3}=1.  \label{27june-3}
\end{equation}
The $d=4$ theory has no non-BPS $Z=0$ ``flat'' directions, and $2$ non-BPS $%
Z\neq 0$ ``flat'' directions \cite
{TT,BFGM1,TT2,Ferrara-Marrani-1,Ferrara-Marrani-2} (in the BH charge
configuration $Q_{0}=\left( p^{0},0,q_{0},0\right) $ they are precisely
given by Eq. (\ref{27june-3})), whereas the $d=5$ theory has no non-BPS
``flat'' directions at all.

Let us start by analyzing the $d=5$ BPS attractors for the BH charge
configuration $Q_{e}=\left( p^{0},0,0,q_{i}\right) $. For the $d=5$ uplift
of the $stu$ model the matrix $a^{ij}$ is diagonal:
\begin{equation}
a_{stu}^{ij}=diag\left( \left( \widehat{\lambda }^{1}\right) ^{2},\left(
\widehat{\lambda }^{2}\right) ^{2},\left( \widehat{\lambda }^{3}\right) ^{2}=%
\frac{1}{\left( \widehat{\lambda }^{1}\right) ^{2}\left( \widehat{\lambda }%
^{2}\right) ^{2}}\right) ,
\end{equation}
and thus the $stu$ \textit{electric} $d=5$ effective BH potential reads
\begin{equation}
V_{5,stu}^{q}=\left( \widehat{\lambda }^{1}\right) ^{2}q_{1}+\left( \widehat{%
\lambda }^{2}\right) ^{2}q_{2}+\frac{q_{3}}{\left( \widehat{\lambda }%
^{1}\right) ^{2}\left( \widehat{\lambda }^{2}\right) ^{2}}.
\end{equation}
Since
\begin{equation}
\frac{\partial V_{5,stu}^{q}}{\partial \widehat{\lambda }^{i}}%
=0,~i=1,2\Leftrightarrow \widehat{\lambda }^{1}\widehat{\lambda }^{2}=\left(
\frac{\left( q_{3}\right) ^{2}}{\left| q_{1}q_{2}\right| }\right) ^{1/3},~~%
\frac{\widehat{\lambda }^{1}}{\widehat{\lambda }^{2}}=\left| \frac{q_{2}}{%
q_{1}}\right| ,
\end{equation}
one obtains
\begin{equation}
\left. \frac{V_{5,stu}^{q}}{3}\right| _{\frac{\partial V_{5,stu}^{q}}{%
\partial \widehat{\lambda }^{i}}=0,~i=1,2}=\left| q_{1}q_{2}q_{3}\right|
^{2/3},  \label{27june-5}
\end{equation}
as anticipated in Eq. (\ref{27june-4}).

The (real) $d=5$, $\mathcal{N}=2$ central charge (in the \textit{electric}
BH charge configuration $Q_{e}$), is defined as
\begin{equation}
Z_{5}^{q}\equiv \widehat{\lambda }^{i}q_{i}.
\end{equation}
Thus, one obtains that the $d=5$ BPS conditions
\begin{equation}
\frac{\partial Z_{5}^{q}}{\partial \widehat{\lambda }^{i}}%
=0,~i=1,2\Leftrightarrow \left( \widehat{\lambda }^{1}\right) ^{2}\widehat{%
\lambda }^{2}=\frac{q_{3}}{q_{1}},~\widehat{\lambda }^{1}\left( \widehat{%
\lambda }^{2}\right) ^{2}=\frac{q_{3}}{q_{2}},  \label{BPS-conds-d=5}
\end{equation}
are solved only by requiring $\frac{q_{3}}{q_{1}}>0$ and $\frac{q_{3}}{q_{2}}%
>0$, \textit{i.e.} for $q_{1}$, $q_{2}$ and $q_{3}$ \textit{all} with the
same sign. Consequently, with no loss of generality we can conclude that the
$d=5$ critical point determining the result (\ref{27june-5}) is BPS if $%
q_{1} $, $q_{2}$, $q_{3}>0$, and it is non-BPS if $q_{1}$, $q_{2}<0$ and $%
q_{3}>0$.\medskip

Now, in order to find the relation with the $d=4$ attractors, one must
compute the (complex) $d=4$, $\mathcal{N}=2$ central charge $Z(z,{\bar z }%
,Q) $ and its covariant derivatives. By recalling the standard definition in
obvious notation, $Z=exp(K/2)(X^{\Lambda}q_{\Lambda}-F_{\Lambda}p^{\Lambda})$%
, going to special coordinates, in the K\"{a}hler gauge ( $X^{0}\equiv 1$),
and exploiting the cubic nature of the holomorphic prepotential $f\left(
z\right) $ given in (\ref{effe}), one gets
\begin{equation}
Z\left( z,\bar{z},Q\right) =e^{K\left( z,\overline{z}\right) /2}\left[
q_{0}+q_{i}z^{i}+p^{0}f\left( z\right) -p^{i}f_{i}\left( z\right) \right]
\equiv e^{K\left( z,\overline{z}\right) /2}W\left( z,Q\right) ,
\end{equation}
where 
$W$ is the $d=4 $, $\mathcal{N}=2$ holomorphic superpotential. In terms of
the real components of the $d=4$ moduli as $z^{i}=x^{i}-i\mathcal{V}^{1/3}%
\widehat{\lambda }^{i}$ one gets the following expressions ($e^{K\left( z,%
\overline{z}\right) /2}=\frac{1}{2\sqrt{2}}\mathcal{V}^{-1/2}$):
\begin{eqnarray}
&&
\begin{array}{l}
Z\left( x^{j},\widehat{\lambda }^{j},\mathcal{V},Q\right) =\frac{1}{2\sqrt{2}%
}\mathcal{V}^{-1/2}W\left( x^{j},\widehat{\lambda }^{j},\mathcal{V},
Q\right) = \\
\\
=\frac{1}{2\sqrt{2}}\mathcal{V}^{-1/2}\left[
\begin{array}{l}
q_{0}+q_{i}x^{i}-i\mathcal{V}^{1/3}q_{i}\widehat{\lambda }^{i}+ \\
\\
+\frac{p^{0}}{6}h-i\frac{p^{0}}{2}\mathcal{V}^{1/3}\widehat{\kappa }%
_{ij}x^{i}x^{j}+ \\
\\
-\frac{p^{0}}{2}\mathcal{V}^{2/3}\widehat{\kappa }_{i}x^{i}+ip^{0}\mathcal{V}%
+ \\
\\
-\frac{p^{i}}{2}h_{i}+ip^{i}\mathcal{V}^{1/3}\widehat{\kappa }_{ij}x^{j}+%
\frac{p^{i}}{2}\mathcal{V}^{2/3}\widehat{\kappa }_{i}
\end{array}
\right] ;
\end{array}
\notag \\
&&  \label{Z-d-SKG}
\end{eqnarray}
\begin{eqnarray}
&&
\begin{array}{l}
D_{i}Z\left( x^{j},\widehat{\lambda }^{j},\mathcal{V},Q\right) =\frac{1}{2%
\sqrt{2}}\mathcal{V}^{-1/2}D_{i}W\left( x^{j},\widehat{\lambda }^{j},%
\mathcal{V},Q\right) = \\
\\
=\frac{1}{2\sqrt{2}}\mathcal{V}^{-1/2}\left\{
\begin{array}{l}
q_{i}+\frac{p^{0}}{2}h_{i}-ip^{0}\mathcal{V}^{1/3}\widehat{\kappa }%
_{ij}x^{j}-\frac{p^{0}}{4}\mathcal{V}^{2/3}\widehat{\kappa }_{i}+ \\
\\
-p^{j}h_{ij}+ip^{j}\mathcal{V}^{1/3}\widehat{\kappa }_{ij}+ \\
\\
-\frac{i}{4}\mathcal{V}^{-1/3}\widehat{\kappa }_{i}\left[
\begin{array}{l}
q_{0}+q_{j}x^{j}-i\mathcal{V}^{1/3}q_{j}\widehat{\lambda }^{j}+ \\
\\
+\frac{p^{0}}{6}h-i\frac{p^{0}}{2}\mathcal{V}^{1/3}\widehat{\kappa }%
_{jk}x^{j}x^{k}-\frac{p^{0}}{2}\mathcal{V}^{2/3}\widehat{\kappa }_{j}x^{j}+
\\
\\
-\frac{p^{j}}{2}h_{j}+ip^{j}\mathcal{V}^{1/3}\widehat{\kappa }_{jk}x^{k}+%
\frac{p^{j}}{2}\mathcal{V}^{2/3}\widehat{\kappa }_{j}
\end{array}
\right]
\end{array}
\right\} .
\end{array}
\notag \\
&&  \label{DiZ-d-SKG}
\end{eqnarray}

Eqs. (\ref{Z-d-SKG}) and (\ref{DiZ-d-SKG}) are general, but for our purposes
we actually need them just for the particular BH charge configuration $%
Q_0=\left( p^{0},0,0,q_{i}\right) $. For this charge vector, Eq. (\ref
{DiZ-d-SKG}) yields that
\begin{equation}
x^{i}=0,\forall i \Longrightarrow Im\left( D_{i}Z\right) =0\, ,
\end{equation}
and moreover, considering the $stu$ model,
\begin{equation}
\frac{q_{3}}{q_{1}},\frac{q_{3}}{q_{2}}>0,p^{0}<0\Longrightarrow Re\left(
D_{i}Z\right) =0\, ,
\end{equation}
with $\widehat{\lambda }^{1}$ and $\widehat{\lambda }^{2}$ given by Eq. (\ref
{BPS-conds-d=5}). Thus, since the $d=5$ BPS configuration has $q_{1}$, $%
q_{2} $ and $q_{3}>0$, it is clear that it will determine a $d=4$ BPS
configuration if the magnetic charge of the Kaluza-Klein vector is negative (%
\textit{i.e.} $p^{0}<0$), and a $d=4$ non-BPS configuration if $p^{0}>0$. In
order to determine whether this latter $d=4$ configuration is non-BPS $Z\neq
0$ or non-BPS $Z=0$, one has to check the $d=4$ central charge. For
vanishing axions, it turns out to be purely imaginary (the asterisk denotes
the evaluation at vanishing axions
treatment ):
\begin{equation}
Z^{\ast }=\frac{i}{2\sqrt{2}}\left[ p^{0}\mathcal{V}^{1/2}-\mathcal{V}%
^{-1/6}q_{i}\widehat{\lambda }^{i}\right] .  \label{27june-6}
\end{equation}

Since at the $d=5$ BPS attractors it holds that $\left( q_{i}\widehat{%
\lambda }^{i}\right) _{d=5,BPS}=3\left( q_{1}q_{2}q_{3}\right) ^{1/3}$, by
using the critical value of $\mathcal{V}$ given by Eq. (\ref{29june-3}) and
considering $p^{0}<0$, one gets that ($p^{0}<0$, $q_{1}>0$, $q_{2}>0$, $%
q_{3}>0$)
\begin{equation}
Z^\ast_{BPS}=-i\sqrt{2}\left( -p^{0}q_{1}q_{2}q_{3}\right) ^{1/4}\neq 0.
\label{27june-8}
\end{equation}
Thus, the $d=4$ $\frac{1}{2}$-BPS attractor determined by the $d=5$ $\frac{1%
}{2}$-BPS attractor configuration $q_{1}$, $q_{2}$, $q_{3}>0$ by considering
$p^{0}<0$ has the following non-vanishing BH entropy:
\begin{equation}
\frac{S_{BH}}{\pi }=V^\ast_{\left( p^{0},q_{i}\right) ,BPS}=\left|
Z^\ast\right| _{BPS}^{2}=2\left( -p^{0}q_{1}q_{2}q_{3}\right) ^{1/2}.
\end{equation}
\medskip

On the other hand, by using the critical value of $\mathcal{V}$ given by Eq.
(\ref{27june-7}) and considering $p^{0}>0$, one gets that the $d=5$ $\frac{1%
}{2}$-BPS attractor configuration $q_{1}$, $q_{2}$, $q_{3}>0$ also
determines a $d=4$ non-BPS attractor with ($p^{0}>0$, $q_{1}>0$, $q_{2}>0$, $%
q_{3}>0$)
\begin{equation}
Z^\ast_{non-BPS}=-\frac{i}{\sqrt{2}}\left( p^{0}q_{1}q_{2}q_{3}\right)
^{1/4}\neq 0,
\end{equation}
yielding a non-vanishing BH entropy
\begin{equation}
\frac{S_{BH}}{\pi }=V^\ast_{\left( p^{0},q_{i}\right) ,non-BPS,Z^\ast\neq
0}=4\left| Z\right| _{non-BPS,Z\neq 0}^{2}=2\left(
p^{0}q_{1}q_{2}q_{3}\right) ^{1/2},  \label{27june-9}
\end{equation}
as found in \cite{BFGM1}.\medskip\ In order to derive Eq. (\ref{27june-9}),
we have used the standard definition of $V$ in terms of central charge and
matter charges,$V=|Z|^2+|D_{i}Z|^2$ and we have computed the purely real
value $D_{i}Z$, also by using the critical value of $\mathcal{V}$ given by
Eq. (\ref{27june-7}).\medskip\

Let us now move to considering the $d=5$ non-BPS attractors for the BH
charge configuration $Q_{e}=\left( p^{0},0,0,q_{i}\right) $. As pointed out
above, in this case one can assume without loss of generality that $q_{1}$, $%
q_{2}<0 $, and $q_{3}>0$ (indeed violating the $d=5$ BPS conditions). This
yields $\left( q_{i}\widehat{\lambda }^{i}\right) _{d=5,non-BPS}=-\left|
q_{1}q_{2}q_{3}\right| ^{1/3}$.

If $p^{0}<0$, one finds that, using the critical value of $\mathcal{V}$
given by Eq. (\ref{27june-7}), the two terms in Eq. (\ref{27june-6})
reciprocally cancel ($p^{0}<0$, $q_{1}<0$, $q_{2}<0$, $q_{3}>0$):
\begin{equation}
Z^\ast=0.
\end{equation}
Thus, the BH charge configuration $p^{0}<0$, $q_{1}$, $q_{2}<0$, and $%
q_{3}>0 $ supports a $d=4$ non-BPS $Z^\ast=0$ attractor.

If $p^{0}>0$ the two terms in Eq. (\ref{27june-6}) sum up into the following
expression ($p^{0}>0$, $q_{1}<0$, $q_{2}<0$, $q^{3}>0$):
\begin{equation}
Z^\ast=\frac{i}{\sqrt{2}}\left( p^{0}q_{1}q_{2}q_{3}\right) ^{1/4}\neq 0.
\end{equation}
Thus, the BH charge configuration $p^{0}>0$, $q_{1}$, $q_{2}<0$, and $%
q_{3}>0 $ supports a $d=4$ non-BPS $Z^\ast\neq 0$ attractor, whose entropy
is given by Eq. (\ref{27june-9}). Thus, concerning the $d=4$
supersymmetry-preserving features, the two BH charge configurations $\left(
p^{0}>0,q_{1}>0,q_{2}>0,q_{3}>0\right) $ (upliftable to $d=5$ BPS) and $%
\left( p^{0}>0,q_{1}<0,q_{2}<0,q_{3}>0\right) $ (upliftable to $d=5$
non-BPS) are \textit{equivalent}.\bigskip

Summarizing, we have found that the BH charge configuration specified by the
charge vector $Q_{e}=\left( p^{0},0,0,q_{i}\right) $ splits into three
\textit{inequivalent} configurations, depending on the signs of the charges.
The critical value of the $d=4$ effective BH potential has the general form
\begin{equation}
V_{\left( p^{0},q_{i}\right) ,crit.}^{\ast }=2\left|
p^{0}q_{1}q_{2}q_{3}\right| ^{1/2},
\end{equation}
but the $\mathcal{N}=2$, $d=4$ central charge correspondingly takes three
different values:
\begin{equation}
\begin{array}{l}
p^{0}<0,q_{1}>0,q_{2}>0,q_{3}>0\Leftrightarrow \frac{1}{2}-BPS:\left|
Z\right| _{\frac{1}{2}-BPS}=\sqrt{2}\left( -p^{0}q_{1}q_{2}q_{3}\right)
^{1/4}; \\
\\
p^{0}<0,q_{1}<0,q_{2}<0,q_{3}>0\Leftrightarrow non-BPS,Z=0:Z_{non-BPS,Z=0}=0;
\\
\\
\left.
\begin{array}{c}
p^{0}>0,q_{1}>0,q_{2}>0,q_{3}>0 \\
\\
p^{0}>0,q_{1}<0,q_{2}<0,q_{3}>0
\end{array}
\right\} \Leftrightarrow non-BPS,Z\neq 0:\left\{
\begin{array}{l}
\left| Z\right| _{non-BPS,Z\neq 0}= \\
\\
=\frac{1}{\sqrt{2}}\left( p^{0}q_{1}q_{2}q_{3}\right) ^{1/4}.
\end{array}
\right.
\end{array}
\end{equation}
\medskip

It is useful to point out that in \cite{FG2} a different normalization for
the $d=5$ (\textit{electric}) effective BH potential was used:
\begin{equation}
V_{5}^{BH}=3V_{5}^{q},  \label{VBH5-V5q}
\end{equation}
due to a different normalization of the $d=5$ vector kinetic matrix: $%
\overset{\circ }{a}_{ij}^{V}=\frac{1}{3}a_{ij}$. As a consequence
\begin{equation}
V_{5,crit.}^{BH}=9\left| q_{1}q_{2}q_{3}\right| ^{2/3},  \label{VBH5-crit}
\end{equation}
and the corresponding $d=5$ entropy is given by Eq. (\ref{SBH5}). Since $%
Z_{5,BPS}^{q}=3\left| q_{1}q_{2}q_{3}\right| ^{1/3}$ and $%
Z_{5,non-BPS}^{q}=\left| q_{1}q_{2}q_{3}\right| ^{1/3}$, one then retrieves
the $d=5$ \textit{sum rules} derived in \cite{FG2}, \textit{i.e.}
\begin{equation}
V_{5,BPS}^{BH}=Z_{5,BPS}^{2},~~V_{5,non-BPS}^{BH}=9Z_{5,non-BPS}^{2},
\end{equation}
which also shows that
\begin{equation}
\frac{3}{2}\left[ g^{xy}\left( \partial _{x}Z_{5}\right) \partial _{y}Z_{5}%
\right] _{non-BPS}=8Z_{5,non-BPS}^{2},
\end{equation}
due to the identity \cite{FG2}
\begin{equation}
V_{5}^{BH}=Z_{5}^{2}+\frac{3}{2}g^{xy}\left( \partial _{x}Z_{5}\right)
\partial _{y}Z_{5},
\end{equation}
where the $d=5$ scalar metric in our convention reads
\begin{equation}
g_{xy}=\frac{1}{2}\left( \partial _{x}\widehat{\lambda }_{i}\right) \left(
\partial _{y}\widehat{\lambda }_{j}\right) a^{ij},
\end{equation}
with $\widehat{\lambda }_{i}\equiv a_{ij}\widehat{\lambda }^{j}$.

\section{\label{Conclusion}\textbf{Conclusion}}

In this study we have related the $d=4$ and $d=5$ entropy formul\ae\ for
special geometry described by a cubic holomorphic prepotential function ($d$%
-geometry), based solely on properties of the general black hole effective
potential (\ref{VBH-d=4-d-SKG}), rather than considering solutions for the
scalar fields. These $d$-geometries are particularly relevant, as they
describe the large volume limit of the $CY_{3}$-compactifications of type
IIA superstrings or, in the $d=5$ uplift, the $CY_{3}$-compactifications of $%
M$-theory \cite{Cadavid-et-al}. $d$-geometries also include homogeneous and
symmetric special geometries, where we have shown that they have even more
interesting properties.

It is now worth commenting about the charge configurations used in our
treatment. For symmetric $d$-geometries the configurations $\left(
p^{0},0,0,q_{i}\right) $ (\textit{electric}, upliftable to $d=5$ extremal
Tangherlini BH with horizon geometry $AdS_{2}\times S^{3}$) and $\left(
0,p^{i},q_{0},0\right) $ (\textit{magnetic}, upliftable to $d=5$ black
string with horizon geometry $AdS_{3}\times S^{2}$) are reciprocally dual.
For the electric configuration $\left( p^{0},0,0,q_{i}\right) $ (and for
vanishing axions) the $d=5$ and $d=4$ effective BH potentials are
respectively denoted by $V_{5}^{q}$ and $V^{\ast }$, and respectively given
by Eqs. (\ref{27june-14}) and (\ref{27june-13}). For the magnetic
configuration $\left( 0,p^{i},q_{0},0\right) $ (and for vanishing axions) \
the $d=5$ and $d=4$ effective BH potentials are respectively denoted by $%
V_{5}^{p}$ and $V^{\ast }$, and respectively given by
\begin{equation}
V_{5}^{p}=a_{ij}p^{i}p^{j}=V_{5}^{p}\left( \widehat{\lambda }%
^{i},p^{j}\right) ;  \label{27june-11}
\end{equation}
\begin{equation}
V^{\ast }\left( \widehat{\lambda }^{i},\mathcal{V},p^{j},q_{0}\right) =\frac{%
1}{2}\left[ \mathcal{V}^{-1}\left( q_{0}\right) ^{2}+\mathcal{V}%
^{1/3}V_{5}^{p}\left( \widehat{\lambda }^{i},p^{j}\right) \right] .
\label{27june-12}
\end{equation}
By comparing Eqs. (\ref{27june-11}) and (\ref{27june-12}) with Eqs. (\ref
{27june-14}) and (\ref{27june-13}), it is easy to realize that such two
pairs of Eqs. are related (in the $stu$ model, thus $i=1,2,3$) by the
transformations
\begin{equation}
\begin{array}{l}
\mathcal{V}\longleftrightarrow \mathcal{V}^{-1}; \\
q_{i}\longleftrightarrow p^{i}; \\
\left| q_{0}\right| \longleftrightarrow \left| p^{0}\right| ; \\
\widehat{\lambda }^{i}\longleftrightarrow \left( \widehat{\lambda }%
^{i}\right) ^{-1}.
\end{array}
\label{mapping}
\end{equation}

The critical values of the potentials given by Eqs. (\ref{27june-11}) and (%
\ref{27june-12}) respectively are
\begin{equation}
\frac{V_{5,crit.}^{p}}{3}=\left| \frac{1}{3!}d_{ijk}p^{i}p^{j}p^{k}\right|
^{2/3}=\left| p^{1}p^{2}p^{3}\right| ^{2/3};  \label{27june-15}
\end{equation}
\begin{equation}
V_{crit.}^{\ast }\left( p^{j},q_{0}\right) =2\left|
q_{0}p^{1}p^{2}p^{3}\right| ^{1/2}.  \label{27june-16}
\end{equation}
Moreover, the $d=5$ volume is related to $V_{5}^{p}$ as follows:
\begin{equation}
\mathcal{V}^{-4/3}=\frac{1}{3\left( q_{0}\right) ^{2}}V_{5}^{p}~,
\label{29june-1}
\end{equation}
yielding the critical value
\begin{equation}
\mathcal{V}_{cr.}^{-4/3}=\frac{1}{\left( q_{0}\right) ^{2}}\left( \frac{1}{3!%
}\left| d_{ijk}p^{i}p^{j}p^{k}\right| \right) ^{2/3}~.  \label{29june-2}
\end{equation}
Notice that Eqs. (\ref{29june-1}) and (\ref{29june-2}) can respectively be
obtained from Eqs. (\ref{27june-7}) and (\ref{29june-3}) by applying the
mapping (\ref{mapping}). The $d=5$ and $d=4$ supersymmetry-preserving
features for the magnetic configuration goes the same way as for the
electric configuration, with the interchanges $p^{0}\longrightarrow -q_{0}$
and $q_{i}\longrightarrow p^{i}$ ($i=1,2,3$).

We remark that Eqs. (\ref{27june-15}) and (\ref{29june-2}) (as well as Eqs. (%
\ref{27june-11}) and (\ref{27june-12})) are valid also for non-symmetric $d$%
-geometries, because they make use of the completely covariant tensor $%
d_{ijk}$ (the same appearing in the holomorphic prepotential). However, let
us stress that for non-symmetric $d$-geometries the (relevant expressions
for the) electric configuration is much more complicated, due to the lack of
a globally constant tensor $d^{ijk}$ (which in the non-symmetric case does
not satisfy the relation (\ref{symmetric-cond})).

It would be intriguing to study the ``flat'' directions of the effective BH
potentials for (symmetric and non-symmetric) $d=5$ (special real) and $d=4$
(special K\"{a}hler) $d$-geometries, and relate the corresponding moduli
spaces \cite{Ferrara-Marrani-2} through the results obtained in the present
work.

Here we limit ourselves to observe that the moduli space of $d=5$ non-BPS
attractors is both a submanifold of the moduli space of the $d=4$ non-BPS $%
Z\neq 0$ attractors (coinciding with the corresponding $d=5$ special real
manifold) and of the moduli space of the $d=4$ non-BPS $Z=0$ attractors.
This non-trivial result, obtained in \cite{Ferrara-Marrani-2}, can also be
understood as an outcome of our analysis of this paper where, for the
so-called electric - and magnetic - charge configurations, we have shown
that the $d=5$ non-BPS attractors can give rise to both classes ($Z\neq 0$
and $Z=0$) of $d=4$ non-BPS attractors. Consequently, the $d=5$ non-BPS
``flat'' directions must be common to both types of $d=4$ non-BPS ``flat''
directions.

It would be interesting to analyze these issues in more detail, and
understand better the ``flat'' directions (and their fate once the quantum
corrections are switched on), as they depend on the background charge
configuration vector $Q$.

\section*{\textbf{Acknowledgments}}

A. C. is grateful to the organizers of the \textit{``School on Attractor
Mechanism 2007''} (Frascati) for providing the lively atmosphere where this
paper was conceived. A.C. also thanks Gianguido Dall'Agata for many fruitful
discussions and interest in this work.

A. M. would like to thank the Department of Physics, Theory Unit Group at
CERN for its kind hospitality during the completion of the present paper.

The work of A.C. and S.F. has been supported in part by European Community
Human Potential Program under contract MRTN-CT-2004-005104 \textit{%
``Constituents, fundamental forces and symmetries of the universe''} and for
S. F. also the contract MRTN-CT-2004-503369 \textit{``The quest for
unification: Theory Confronts Experiments''}, and by D.O.E. grant
DE-FG03-91ER40662, Task C.

The work of A.M. has been supported by a Junior Grant of the \textit{%
``Enrico Fermi''} Center, Rome, in association with INFN Frascati National
Laboratories.

\end{document}